\documentclass[twocolumn,showpacs,preprintnumbers,amsmath,amssymb]{revtex4}
\usepackage{graphicx}
\usepackage{dcolumn}
\usepackage{bm}

\begin{document}
\def\d{{\rm d}}
\def\Epos{E_{\rm pos}}
\def\ap{\approx}
\def\eff{{\rm eff}}
\def\L{{\cal L}}
\newcommand{\vev}[1]{\langle {#1}\rangle}
\newcommand{\CL}   {C.L.}
\newcommand{\dof}  {d.o.f.}
\newcommand{\eVq}  {\text{eV}^2}
\newcommand{\Sol}  {\textsc{sol}}
\newcommand{\SlKm} {\textsc{sol+kam}}
\newcommand{\Atm}  {\textsc{atm}}
\newcommand{\Chooz}{\textsc{chooz}}
\newcommand{\Dms}  {\Delta m^2_\Sol}
\newcommand{\Dma}  {\Delta m^2_\Atm}
\newcommand{\Dcq}  {\Delta\chi^2}
\newcommand{\nbb}{$\beta\beta_{0\nu}$ }
\def\e6{E(6)}
\def\10{SO(10)}
\def\21{SU(2) $\otimes$ U(1) }
\def\321{$\mathrm{SU(3) \otimes SU(2) \otimes U(1)}$ }
\def\lr{SU(2)$_L \otimes$ SU(2)$_R \otimes$ U(1)}
\def\422{SU(4) $\otimes$ SU(2) $\otimes$ SU(2)}
\newcommand{\AHEP}{%
  AHEP Group, Instituto de F\'{\i}sica Corpuscular --
  C.S.I.C./Universitat de Val{\`e}ncia \\
  Edificio Institutos de Paterna, Apt 22085, E--46071 Valencia, Spain}
\newcommand{\RIVER}{%
Physics Department, University of California, Riverside, CA 92521, USA, 
and\\ Institute for Particle Physics Phenomenology, University of Durham, 
Durham, DH1 3LE, UK}
\def\roughly#1{\mathrel{\raise.3ex\hbox{$#1$\kern-.75em
      \lower1ex\hbox{$\sim$}}}} \def\lsim{\roughly<}
\def\gsim{\roughly>}
\def\ltap{\raisebox{-.4ex}{\rlap{$\sim$}} \raisebox{.4ex}{$<$}}
\def\gtap{\raisebox{-.4ex}{\rlap{$\sim$}} \raisebox{.4ex}{$>$}}
\def\lsim{\raise0.3ex\hbox{$\;<$\kern-0.75em\raise-1.1ex\hbox{$\sim\;$}}}
\def\gsim{\raise0.3ex\hbox{$\;>$\kern-0.75em\raise-1.1ex\hbox{$\sim\;$}}}

\preprint{IFIC/05-32}
\title{Predicting Neutrinoless Double Beta Decay}
\date{\today}
\author{M. Hirsch}\email{Martin.Hirsch@ific.uv.es}
\affiliation{\AHEP}
\author{Ernest Ma}\email{ma@phyun8.ucr.edu}
\affiliation{\RIVER}
\author{A. Villanova del Moral}\email{Albert.Villanova@ific.uv.es}
\affiliation{\AHEP}
\author{J. W. F. Valle}\email{valle@ific.uv.es}
\affiliation{\AHEP}
\begin{abstract}
  
  We give predictions for the neutrinoless double beta decay rate in a
  simple variant of the $A_4$ family symmetry model. We
  show that there is a {\sl lower} bound for the \nbb amplitude even
  in the case of normal hierarchical neutrino masses, corresponding to
  an effective mass parameter $ \left| m_{ee} \right| \geq 0.17
  {\sqrt{\Dma}}$.  This result holds both for the CP
  conserving and CP violating cases. In the latter case we show
  explicitly that the lower bound on $\left| m_{ee} \right|$ is
  sensitive to the value of the Majorana phase.  We conclude therefore 
  that in our scheme, \nbb may be accessible to the next generation of
  high sensitivity experiments.

 \end{abstract}
\maketitle


Are neutrinos their own antiparticles?  This is one of the most basic
current unknowns in neutrino physics.  Of all possible manifestations
of neutrino masses, neutrinoless double beta decay (\nbb, for short)
offers -- to date -- the only potentially viable way to answer this
question. If \nbb exists, then neutrino masses are Majorana in nature, 
irrespective of their ultimate origin~\cite{schechter:1982bd}.
Together with cosmology~\cite{Hannestad:2004bq} and direct kinematical 
searches in tritium decay~\cite{Drexlin:2005zt}, \nbb offers one of the 
three main complementary ways to probe the absolute scale of neutrino masses.

Current experimental limits from \nbb on the effective Majorana mass
of the neutrino $m_{ee}$ are of order $m_{ee} \le 0.3-1$ eV
\cite{Baudis:1999xd,Aalseth:2002rf}. We note that a claim for a finite
\nbb rate has been published in \cite{Klapdor-Kleingrothaus:2004ge},
but this has so far not been confirmed by any other experiment. A
recent proposal \cite{Abt:2004yk} aims explicitly at testing the
half-life range suggested in \cite{Klapdor-Kleingrothaus:2004ge}.
However, future \nbb experiments may be able to reach down to much lower
mass scales.  Many experiments sensitive to $m_{ee} \simeq 0.05$ eV
have already been discussed; see for example \cite{supernemo,Ardito:2005ar}.
In the longer-term future, even $m_{ee} \simeq 0.01$ eV
\cite{Gaitskell:2003zr,Akimov:2005mq} does not seem impossible.

The historic confirmation of neutrino oscillations over the last few
years~\cite{Maltoni:2004ei}, together with some basic theory, suggests 
that \nbb is expected, although in general no lower bound on the
magnitude of the expected effect can be given.
Theoretical input is therefore needed. Currently the origin of neutrino
masses is completely unknown. The basic dimension--five operator which
leads to neutrino masses~\cite{Weinberg:1980bf} can arise from a
variety of mechanisms characterized by vastly different scales.
Alternatives include the seesaw
mechanism~\cite{Minkowski:1977sc,schechter:1980gr,mohapatra:1981yp}
and low--energy $R-$parity violating supersymmetry~\cite{Hirsch:2004he}. 
In neither case is it possible, in
general, to establish a lower bound on the magnitude of the \nbb
rate.

Here we consider a simple phenomenological model based on a new
realization of the $A_4$ family symmetry~\cite{Ma:2001dn,babu:2002dz,A4misc}
in which no \321 singlet neutrinos are introduced. Instead, the small
neutrino masses arise from the small induced vacuum expectation values
(VEVs) generated for the neutral components of triplet Higgs
bosons~\cite{schechter:1980gr,masarkar}, transforming
nontrivially under the $A_4$ family symmetry. The lepton and Higgs
particle content and their transformation properties under $A_4$ and
\21 are specified in Table I.
\begin{table}[h!]
\label{table:pc}
\renewcommand{\arraystretch}{1.7}
\begin{tabular}{|c||c||c||c|c|c||c|c|c||c|}\hline
\textrm{Fields} & $L$ & $l^c$ & $\phi_1$ & $\phi_2$ & $\phi_3$ & $\eta_1$ & $\eta_2$ & $\eta_3$ & $\xi$\\
\hline\hline
$A_4$ & $\mathbf{3}$ & $\mathbf{3}$ & $\mathbf{1}$ & $\mathbf{1'}$ & $\mathbf{1''}$ & $\mathbf{1}$ & $\mathbf{1'}$ & $\mathbf{1''}$ & $\mathbf{3}$\\
\hline \hline
$SU(2)_L$ & $\mathbf{2}$ & $\mathbf{1}$ & \multicolumn{3}{c||}{$\mathbf{2}$} & \multicolumn{3}{c||}{$\mathbf{3}$} & $\mathbf{3}$\\
\hline
$Y$ & --1 & 2  & \multicolumn{3}{c||}{--1} &  \multicolumn{3}{c||}{2} & 2 \\
\hline
  \end{tabular}
  \caption{Lepton and scalar boson quantum numbers}
\end{table}
With these transformation properties, the charged lepton mass matrix
is already diagonal in the flavor basis, with
\begin{eqnarray*}
 m_e     &  = & h_1v_1+h_2v_2+h_3v_3\\
 m_{\mu} &  = & h_1v_1+\omega h_2v_2+\omega^2h_3v_3\\
 m_{\tau}&  = & h_1v_1+\omega^2 h_2v_2+\omega h_3v_3
\end{eqnarray*}
where $h_i$ are charged lepton Yukawa couplings, $v_i =\vev{\phi_i^0}$
and $\omega$ is a complex cubic root of unity satisfying $1+\omega+\omega^2=0$.
The neutrino mass matrix is then of the form
\begin{equation}
M_{\nu}=\left(
\begin{array}{ccc}
a+b+c & f & e \\
f & a+\omega b+\omega^2c & d \\
e & d & a+\omega^2 b+\omega c 
\end{array}
\right)
\end{equation}
where the only non-diagonal entries are those of the $A_4$ triplet $\xi$.
Here we have defined
\begin{equation}
\begin{aligned}
a&=\lambda_1 \vev{\eta_1^0} \\
b&=\lambda_2 \vev{\eta_2^0} \\
c&=\lambda_3 \vev{\eta_3^0} 
\end{aligned}\qquad\qquad\qquad
\begin{aligned}
d&=\kappa \vev{\xi_1^0}\\
e&=\kappa \vev{\xi_2^0}\\
f&=\kappa \vev{\xi_3^0}
\end{aligned}
\end{equation}
where $\lambda_i, \: \kappa$ and $\vev{\eta_i^0}, \: \vev{\xi_i^0}$ are 
triplet Yukawa couplings, and VEVs, respectively. Let us further assume that
the conditions $b=c$, and $d=e=f$ hold.  Whereas the former is an 
{\it ad hoc} 
assumption, the latter can be maintained naturally because of a residual 
$Z_3$ symmetry. Note that $M_\nu$ has a very remarkable (and possibly unique) 
property here in that each entry is renormalized by the charged-lepton Yukawa 
couplings in the same way, i.e. $|h_1|^2+|h_2|^2+|h_3|^2$, instead of being 
proportional to $m_i^2+m_j^2$ as in the Standard Model.\\

It is easy to see that in the above limit we have the prediction
$$
\theta_{23} =\pi/4, \: \: \: \: \theta_{13} =0$$
which matches well
with the neutrino oscillation data~\cite{Maltoni:2004ei}.  Furthermore
it can be shown that, in the limit where the solar mass splitting is
neglected, $b$ and $d$ can be made real, so that the atmospheric
neutrino mass splitting takes on a very simple form
\begin{equation}
\label{6bd}
\Delta m^2_{32}  = 6bd \equiv \Dma
\end{equation}
The solar neutrino mass splitting $\Dms \ll \Dma$ can be expressed as
\begin{equation}
\label{eq:squares}
\Delta m^2_{21}= \sqrt{T_1^2+T_2^2+T_3^2} \equiv \Dms
\end{equation}
where
\begin{align}
T_1&\equiv 6\sqrt{2}|b||d| \sin(\phi_2)\\
T_2&\equiv 2\sqrt{2}|d|\Big(2|a|\cos(\phi_1)+|b| \cos(\phi_2)+|d|\Big)\\
T_3
&=-3|b|^2+|d|^2-6|a||b|\cos(\phi_1+\phi_2)\\ 
& + 2|a||d|\cos(\phi_1)-2|b||d|\cos(\phi_2) \nonumber
\end{align}
with $\phi_1 \equiv \phi_a-\phi_d$ and $\phi_2 \equiv \phi_d-\phi_b$, 
where $a=|a|\exp(i\phi_a)$, $b=|b|\exp(i\phi_b)$ and
$d=|d|\exp(i\phi_d)$.
The condition in Eq.~(\ref{eq:squares}) leads to three inequalities
$$
|T_i| \le \Dms$$
which, normalized by $\Dma$, can be expressed in terms of the small 
parameter $\alpha \equiv \Dms/\Dma$ as
\begin{equation}
\label{eq:boundary2-1}
\sqrt{2}|\sin(\phi_2)|\le \alpha
\end{equation}
\begin{equation}
\label{eq:boundary2-2}
\frac{\sqrt{2}}{3|b|}\Big|2|a|\cos(\phi_1)+|b|\cos(\phi_2)+|d|\Big|\le \alpha
\end{equation}
\begin{align}
\label{eq:boundary2-3}
\frac{1}{6|b||d|}\Big|-3|b|^2+|d|^2-6|a||b| \cos(\phi_1+\phi_2)\\
+2|a||d|\cos(\phi_1)-2|b||d|\cos(\phi_2)\Big|\le\alpha \nonumber
\end{align}
where the current allowed values of $\alpha$ are
shown in Fig.13 of Ref.~\cite{Maltoni:2004ei}.\\

\begin{figure}[htbp]
  \centering
\includegraphics[clip,height=4cm,width=.7\linewidth]{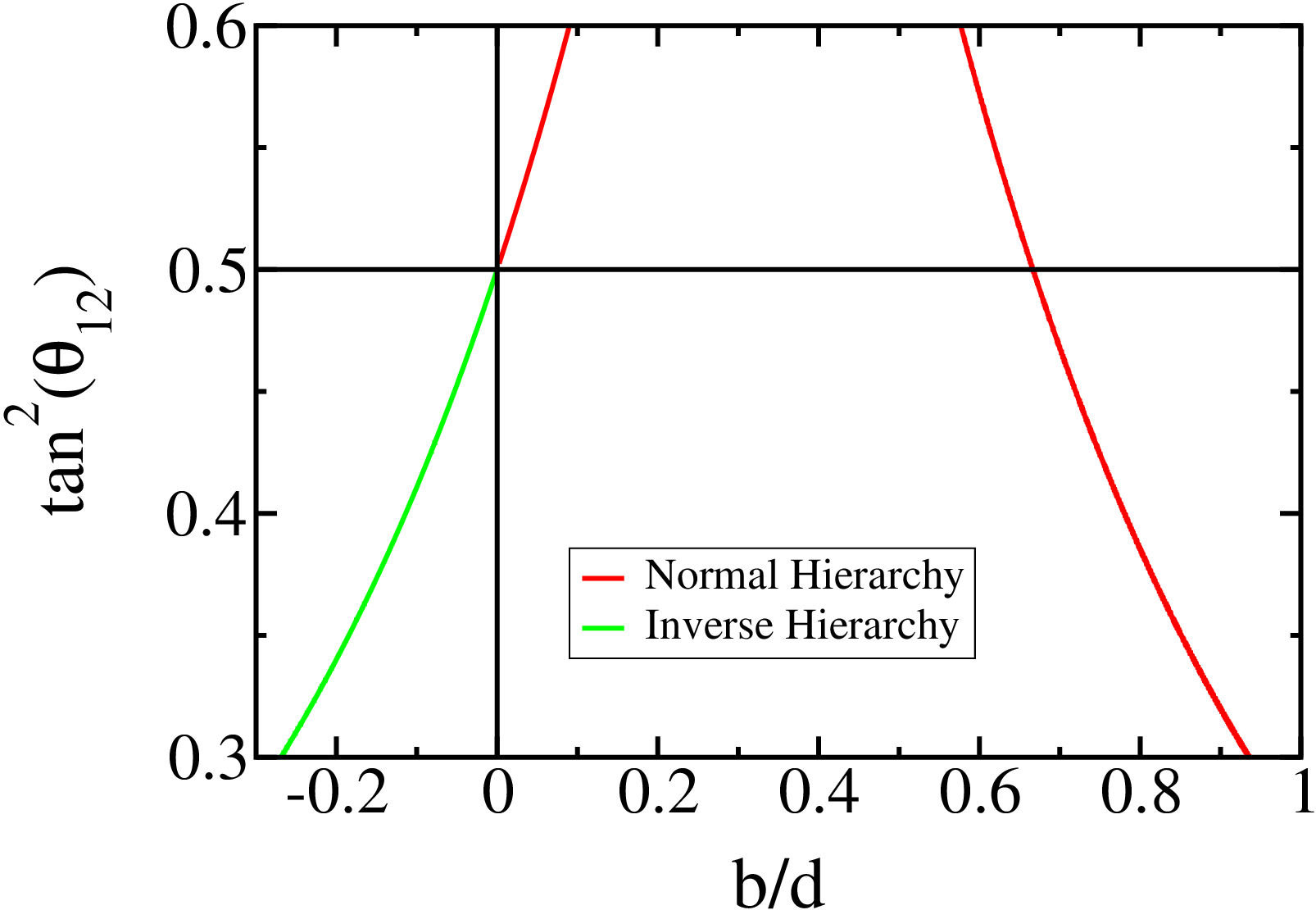}
  \caption{Solar mixing angle, $\tan^2\theta_{12}$, vs. $b/d$.}
  \label{fig:2-branches}
\end{figure}

As for the solar mixing angle, it is given here by
\begin{align}
\label{eq:tan2sol}
t_{2S} \equiv \tan(2\theta_{12})& = \frac{2\sqrt{2}d}{3b-d}
\end{align}
which reduces to 
\begin{equation}
\tan^2\theta_{12}= 1/2
\end{equation}
in two ways, namely,
\begin{align}
\label{eq:b-sol}
b&=0, & b&=2d/3
\end{align}
Current fits of solar, reactor, atmospheric and accelerator neutrino
oscillation data lead to a best fit point to the solar mixing angle
for which $\tan^2\theta_{12}$ is slightly less than 1/2.  Performing a
series expansion of $\tan^2\theta_{12}$ around the two solutions in
Eq.~(\ref{eq:b-sol}), we get
\begin{align}
\tan^2\theta_{12}&\simeq\frac{1}{2}+\frac{b}{d}\\
\tan^2\theta_{12}&\simeq\frac{1}{2}-\frac{1}{d}\left(b-\frac{2}{3}d\right)
\end{align}
These two branches are depicted in Fig.~\ref{fig:2-branches}.
The positive horizontal axis corresponds to a normal hierarchy (NH)
neutrino spectrum, while the negative part corresponds to inverse
hierarchy (IH), as depicted in Fig.~\ref{fig:2-branches}. This behavior is 
also recognized in the model of~\cite{cfm}.\\

\begin{figure}[htbp]
  \centering
  \includegraphics[clip,height=4cm,width=0.47\linewidth]{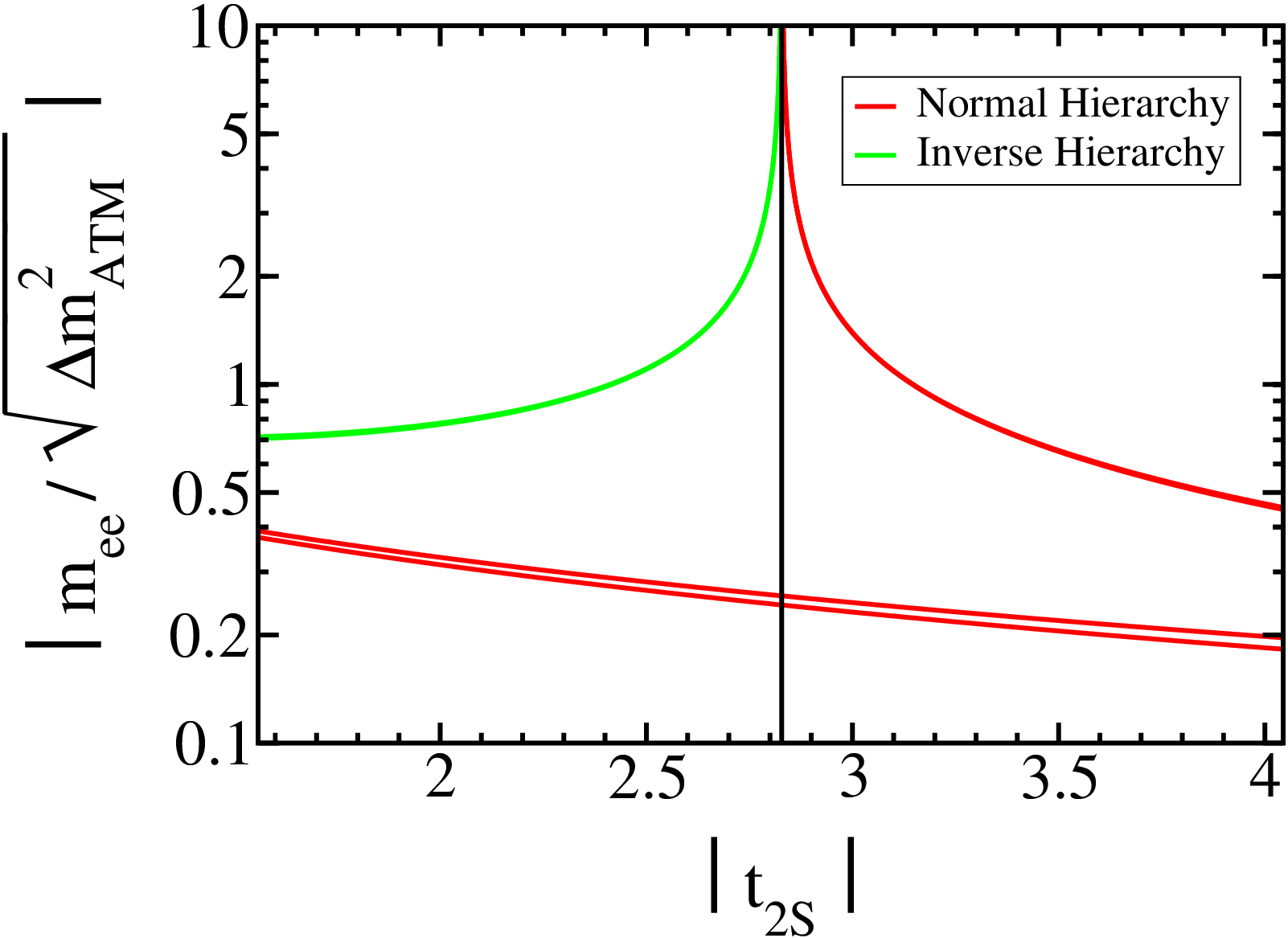} 
  \includegraphics[clip,height=4cm,width=0.47\linewidth]{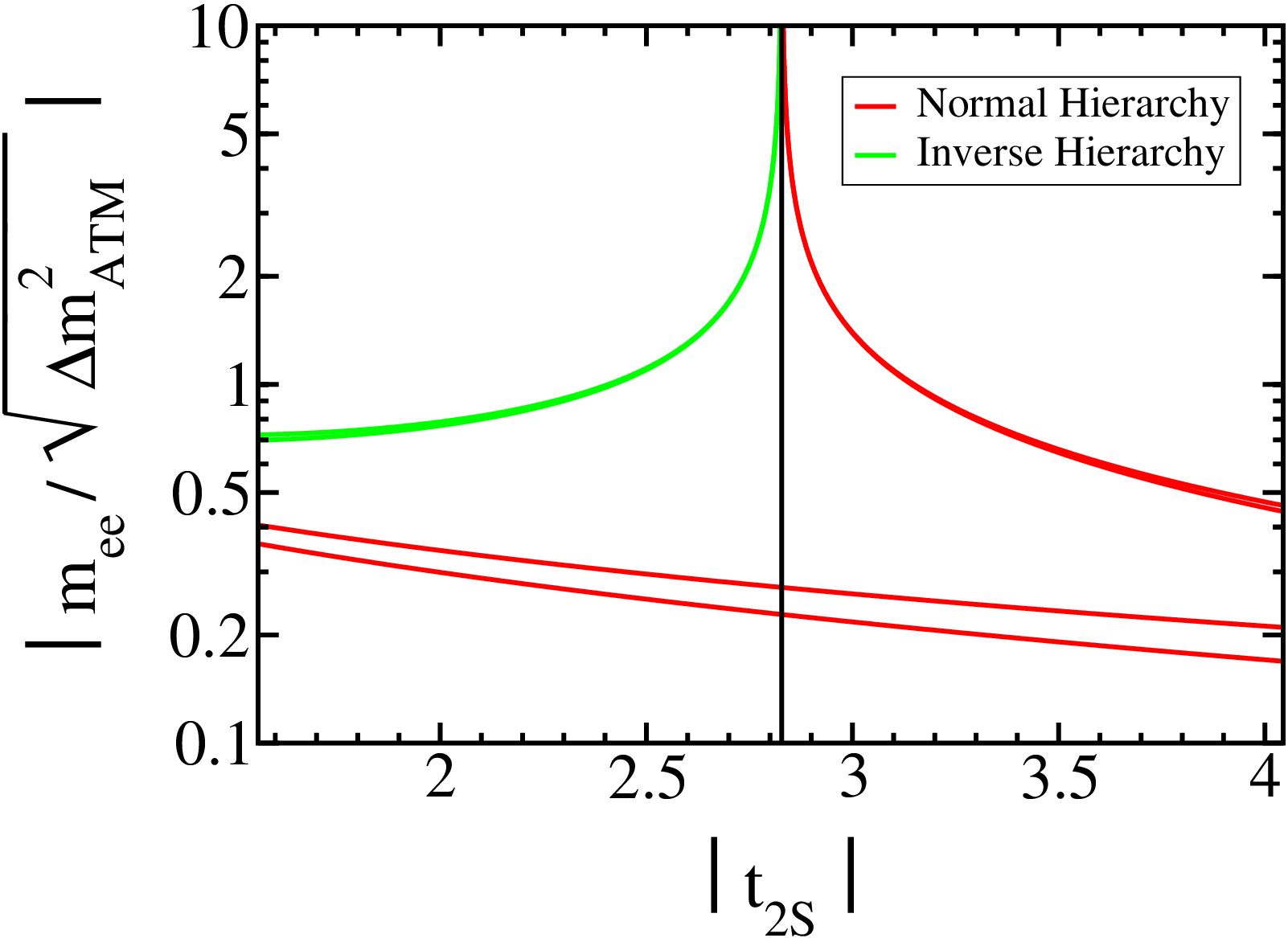} 
  \caption{$|m_{ee}/\sqrt{\Dma}|$ vs. $|t_{2S}|$ according 
    to Eq.~(\ref{eq:mee}).  The bound depends slightly on the value of
    $\alpha$: the left panel corresponds to $\alpha=0.022$ and the
    right one to $\alpha=0.065$.  The dark (red) lines correspond to
    normal hierarchy, while the grey (green) line is for the inverse
    hierarchy case.  The vertical line corresponds to the horizontal
    line in Fig.~\ref{fig:2-branches} at $\tan^2\theta_{12}=1/2$.}
  \label{fig:mee}
\end{figure}

We now turn to neutrinoless double beta decay. The neutrino--exchange
amplitude for this process is simply given by
\begin{equation}
\label{eq:mee-general}
\vev{ m_{\nu}} = m_{ee} =a+2b
\end{equation}
In the case of real parameters, we can solve the following system of
equations
\begin{equation*}
\left.
\begin{gathered}
\Dms \equiv\Delta m_{21}^2=|2a+b+d|\sqrt{(d-3b)^2+8d^2}\\
\Dma \equiv\Delta m_{32}^2=6bd\\
t_{2S}\equiv\tan(2\theta_{12})=\frac{2\sqrt{2}d}{3b-d}
\end{gathered}
\quad\right\}
\end{equation*}
and express the parameters $a$, $b$ and $d$ in terms of experimentally
measurable ones $\Dms$, $\Dma$ and $t_{2S}$. Substituting in
Eq.~(\ref{eq:mee-general}) we can therefore express $\vev{ m_{\nu}}$
in terms of these measured observables. We then obtain, up to an overall
sign,
\begin{align}
\label{eq:mee}
\frac{m_{ee}}{\sqrt{\Dma}} = \textrm{Sign}[\Dma]\frac{1}{\sqrt{2\sqrt{2}t_{2S}+t_{2S}^2}}  \\ \nonumber
\pm \textrm{Sign}[2\sqrt{2}t_{2S}+t_{2S}^2]\frac{ \alpha \sqrt{2\sqrt{2}t_{2S}+t_{2S}^2}}{4\sqrt{1+t_{2S}^2}}
\end{align}
The calculated values of $|m_{ee}/\sqrt{\Dma}|$ as functions of
$t_{2S}$ according to Eq.~(\ref{eq:mee}) are shown in
Fig.~\ref{fig:mee}.
It can be seen that given the currently allowed experimental $3\sigma$
range 
\begin{equation}
\tan^2\theta_{12}\in[0.30,\, 0.61]
\end{equation}
we can set lower bounds for $|m_{ee}/\sqrt{\Dma}|$:
\begin{align}
0.17 & < |m_{ee}/\sqrt{\Dma}|\qquad\textrm{for NH}\\
0.70 & < |m_{ee}/\sqrt{\Dma}|\qquad\textrm{for IH}
\end{align}
Note that the lower bound on \nbb for the NH case is especially
relevant, as it is absent in the generic case.  It results from this
specific realization of the $A_4$ symmetry and is also in contrast with
previous $A_4$--based models that led to quasi--degenerate
neutrinos~\cite{babu:2002dz,Hirsch:2003dr}.

Moreover, should future precision experiments narrow down the
experimental range for $\tan^2\theta_{12}$ then we might be able to
distinguish between both neutrino mass hierarchies.
For example, if $\tan^2\theta_{12} \leq 1/2$ could ever be
established, then we would have
\begin{align}
0.23 & < |m_{ee}/\sqrt{\Dma}|<0.41\qquad\textrm{for NH}\\
0.70 & < |m_{ee}/\sqrt{\Dma}|\qquad\qquad\quad\textrm{for IH}
\end{align}
It can also be seen that, up to order $\alpha$ corrections, $m_{ee}$ can
only be zero if $\tan^2\theta_{12}=1$, now strongly rejected
experimentally.
Note that the two solutions in Fig.~\ref{fig:mee} correspond to the
two branches depicted in Fig.~\ref{fig:2-branches}.  One can see that
only in the branch corresponding to the solution $b=0$ and for values
of $\tan^2\theta_{12}$ that are less than 1/2, there is a relative minus 
sign between $b$ and $d$, which is the condition for inverse
hierarchy, as can be seen from Eq.~(\ref{6bd}).

\begin{figure}[htbp]
  \centering
  \includegraphics[clip,height=5cm,width=0.7\linewidth]{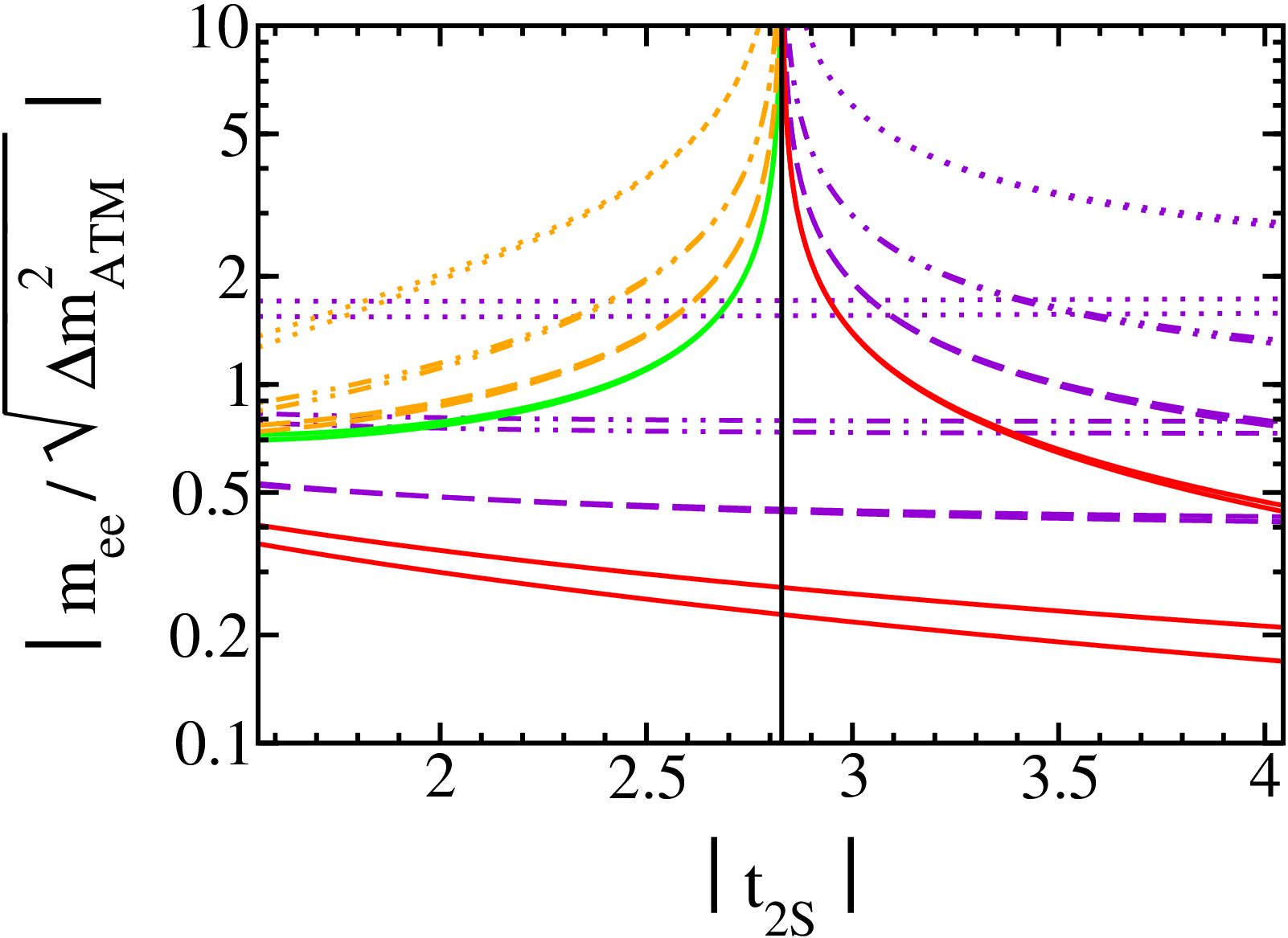} 
  \caption{$|m_{ee}/\sqrt{\Dma}|$ vs. $|t_{2S}|$ for  $\alpha=0.065$ 
    when complex parameters are allowed.}
  \label{fig:mee-complex}
\end{figure}

In the general case of complex parameters, lower bounds on 
$|m_{ee}/\sqrt{\Dma}|$ can also be established for each hierarchy.
This task is simplified greatly by taking into account 
the reliable approximations
$\sin\phi_2 \simeq 0$, and $2|a|cos\phi_1+|b|cos\phi_2+|d| \simeq 0$. 
From Fig.~\ref{fig:mee-complex}, we see that in the complex case, 
lower bounds on $|m_{ee}/\sqrt{\Dma}|$ are indeed also established
for each hierarchy.  By comparing Fig.~\ref{fig:mee-complex} with the
right panel in Fig.~\ref{fig:mee} we find that these lower bounds are
in fact exactly the same as obtained in the real case. The robustness
of these bounds is easily understood. It follows from the fact that
the maximum degree of destructive interference between the three
neutrino--exchange contributions occurs in the real case with
appropriate CP parities~\cite{wolfenstein:1981rk}.

The left-rising lines correspond to inverse hierarchy, while all
others refer to the case of normal hierarchy. The lower--lying pairs
of solid lines correspond to the lower bounds already discussed in
Fig.~\ref{fig:mee}.
Again, the vertical line corresponds to $\tan^2\theta_{12}=1/2$.
Finally, the remaining lines correspond to non-zero values of the
relevant CP-violating phase varying $\cos(\phi_1)$ over the range [0,1] in
equally--spaced steps.  It is conceptually interesting to note that
this phase is ``Majorana type''~\cite{schechter:1980gr,doi:1981yb}, as
we are still considering the case $\theta_{13}=0$.

\begin{figure}[htbp]
  \centering
  \includegraphics[clip,height=4.5cm,width=0.7\linewidth]{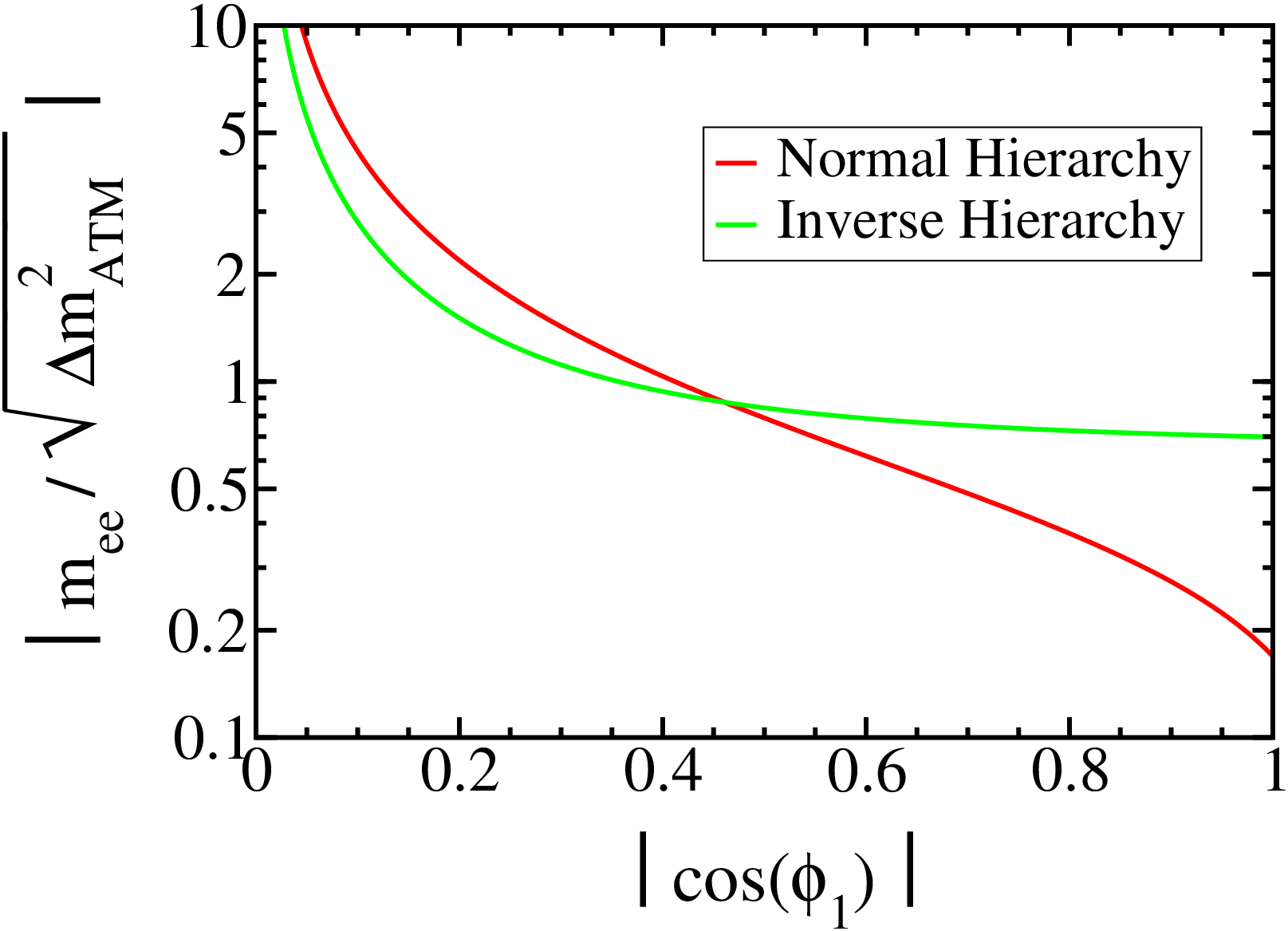}
  \caption{
    Lower bound on $|m_{ee}/\sqrt{\Dma}|$ vs. $|\cos(\phi_1)|$, given
    current $t_{2S}$ uncertainty.  The lines in dark (red) and grey (green)
    correspond to normal and inverse hierarchy respectively.}
  \label{fig:mee-complex-c1}
\end{figure}

Last, but not least, we find it very interesting also that the lower
bound on $|m_{ee}/\sqrt{\Dma}|$ which we have obtained depends on the
value of the Majorana violating phase $|\cos(\phi_1)|$. We can see
from Fig.~\ref{fig:mee-complex-c1} that the lower bounds on
$|m_{ee}/\sqrt{\Dma}|$ for each hierarchy become weaker for
$\cos(\phi_1)= 1$.
Finally, we mention that, even though for simplicity we have focused
on the case $b=c$, all results can be generalized to the case $b\neq
c$, in which case $\theta_{13}$ is allowed to be nonzero.

In summary, we have given predictions for the neutrinoless double beta
decay rate in a simple hierarchical variant of the $A_4$ family symmetry
model. We showed that there is a {\sl lower} bound for the \nbb
amplitude even in the case of normal hierarchical neutrino masses.  We
have seen that the bound is robust as it holds irrespective of whether
CP is conserved or not. In the latter case we show explicitly how the
lower bound on $\left| m_{ee} \right|$ is sensitive to the value of
the Majorana phase. Our scheme suggests that neutrinoless double beta
decay may be within reach of the next generation of high sensitivity
experiments.

This work was supported by the Spanish grant BFM2002-00345 and by the
EC Human Potential Programme RTN network MRTN-CT-2004-503369.  M. H.
is supported by the Ramon y Cajal programme. A. V. M. is supported by
Generalitat Valenciana. The work of E.M. was supported in part by the 
U.S.~Department of Energy under Grant No.~DE-FG03-94ER40837.


\end{document}